\documentclass[conference]{IEEEtran}

\usepackage{graphicx}
\usepackage[utf8]{inputenc} 
\usepackage{amsmath,amssymb,cite,url}
\usepackage{color}
\usepackage{lineno}
\usepackage{latexsym}
\usepackage{color}
\usepackage{siunitx}
\usepackage{amsfonts,amssymb}
\usepackage{tabularx}
\usepackage{booktabs}
\usepackage{multirow}
\usepackage{bm}
\usepackage{wasysym}
\usepackage{subcaption}

\newcommand{\tabref}[1]{{Table \ref{#1}}}
\newcommand{\secref}[1]{{Section \ref{#1}}}

\newcommand{\figref}[1]{{Figure \ref{#1}}}

\begin{document}
%
\title{IDMT-Traffic: An Open Benchmark Dataset for Acoustic Traffic Monitoring Research}


\makeatletter
\newcommand{\linebreakand}{%
  \end{@IEEEauthorhalign}
  \hfill\mbox{}\par
  \mbox{}\hfill\begin{@IEEEauthorhalign}
}
\makeatother

\author{\IEEEauthorblockN{Jakob Abeßer}
\IEEEauthorblockA{\textit{Semantic Music Technologies} \\
\textit{Fraunhofer IDMT}\\
Ilmenau, Germany \\
jakob.abesser@idmt.fraunhofer.de}
\and
\IEEEauthorblockN{Saichand Gourishetti}
\IEEEauthorblockA{\textit{Industrial Media Applications} \\
\textit{Fraunhofer IDMT}\\
Ilmenau, Germany}
\and
\IEEEauthorblockN{András Kátai}
\IEEEauthorblockA{\textit{Industrial Media Applications} \\
\textit{Fraunhofer IDMT}\\
Ilmenau, Germany}
\linebreakand  
\IEEEauthorblockN{Tobias Clauß}
\IEEEauthorblockA{\textit{Industrial Media Applications} \\
\textit{Fraunhofer IDMT}\\
Ilmenau, Germany}
\and
\IEEEauthorblockN{Prachi Sharma}
\IEEEauthorblockA{\textit{Semantic Music Technologies} \\
\textit{Fraunhofer IDMT}\\
Ilmenau, Germany}
\and
\IEEEauthorblockN{Judith Liebetrau}
\IEEEauthorblockA{\textit{Industrial Media Applications} \\
\textit{Fraunhofer IDMT}\\
Ilmenau, Germany}

}

\maketitle

\begin{abstract}
In many urban areas, traffic load and noise pollution are constantly increasing. Automated systems for traffic monitoring are promising countermeasures, which allow to systematically quantify and predict local traffic flow in order to to support municipal traffic planning decisions.
In this paper, we present a novel open benchmark dataset, containing 2.5 hours of stereo audio recordings of 4718 vehicle passing events captured with both high-quality sE8 and medium-quality MEMS microphones. This dataset is well suited to evaluate the use-case of deploying audio classification algorithms to embedded sensor devices with restricted microphone quality and hardware processing power.
In addition, this paper provides a detailed review of recent acoustic traffic monitoring (ATM) algorithms as well as the results of two benchmark experiments on vehicle type classification and direction of movement estimation using four state-of-the-art convolutional neural network architectures.

\end{abstract}


%
\IEEEpeerreviewmaketitle

\section{Introduction}
\label{sec:introduction}

A world-wide rise in population and a steady urbanization trend causes people to move from rural areas to bigger cities.
With more and more active vehicles, travelling times increase and so do noise and air pollution levels.
Intelligent transportation systems (ITS) are effective countermeasures to reduce and optimize traffic flow by adapting to local traffic situations. 
In the past decade, several automatic methods for traffic monitoring were developed for application scenarios such as controlling traffic light cycles, traffic accident detection, logistics monitoring, and other smart city application.

Traffic monitoring systems use various sensor modalities to measure traffic flow ranging from camera sensors for visual object detection and tracking, magnetic loop sensors for counting passing vehicles, to measurement systems based on radio waves (Radar) and light waves (Lidar).
While such systems can be installed as distributed sensor networks to cover large areas, installation and maintenance costs are often high.
Acoustic traffic monitoring (ATM) provides a cheaper alternative for non-intrusive traffic measurements and is the sole focus of this paper.

This paper has three main contributions. First, we present a compact state-of-the-art review of recent ATM systems.
As a second contribution, we introduce the \texttt{IDMT-Traffic} dataset, a novel dataset for traffic monitoring that includes around 2.5 hours of
multi-microphone audio recordings with 4718 annotated passing vehicles. The dataset is intended as public benchmark to further stimulate research on traffic monitoring.
Finally, we present the results of two benchmark experiments for vehicle type classification and direction of movement estimation using four different convolutional neural networks (CNNs) architectures.

This paper is structured as follows: We first review recent ATM algorithms in \secref{sec:related_work} before \secref{sec:dataset} describes the \texttt{IDMT-Traffic} dataset in details.
Then, \secref{sec:experiments} discusses the experimental procedure and the results of the two benchmark experiments. Finally, \secref{sec:conclusion} concludes this work.

\section{Related Work}
\label{sec:related_work}



The audible sound on a road is a combination of several sound sources such as the engines, the exhausts, the wheels and air turbulence, which occur when vehicles pass by \cite{Borkar:2013:NFC:ICFS}. As a reasonable way to break down this complex audio analysis task, researches approached traffic monitoring from different perspectives.
In this section, we categorize existing ATM algorithms based on the approaches for audio data acquisition and statistical modeling.

Moving sound sources such as vehicles can be detected based on their emitted sound if they are recorded with at least two microphones.
Therefore, most ATM methods analyze either \textit{stereo audio signals} \cite{Bhandarkar:2014:VMC:ICCICN, Borkar:2013:NFC:ICFS} \cite{Warghade:2017:RAcoustics:ICCCUBEA} \cite{Tyagi:2012:VTDSE:IEEETTS} \cite{Gatto:2020:MLTCD:IEEETTS} \cite{Djukanovic:2020:Traffic:IVS} or  \textit{multi-channel audio recordings} \cite{Chen:2001:DSFM:IEVT} \cite{Na:2015:ADNI:IEEEIC} \cite{Barbagli:2012:ASNVTM:ICAVSTA} \cite{Ishida:2016:DTW:ITSWC}, which are recorded with 
microphone arrays \cite{Chen:2001:DSFM:IEVT, Na:2015:ADNI:IEEEIC, Barbagli:2012:ASNVTM:ICAVSTA}.
Microphones are commonly integrated into sensor units, which are either placed at the roadside \cite{Warghade:2017:RAcoustics:ICCCUBEA} \cite{Tyagi:2012:VTDSE:IEEETTS} \cite{Na:2015:ADNI:IEEEIC} \cite{Bhandarkar:2014:VMC:ICCICN} or mounted on light poles at a height between 0.5 and 3 meters \cite{Borkar:2013:NFC:ICFS} \cite{Chen:2001:DSFM:IEVT} \cite{Djukanovic:2020:Traffic:IVS}. 


\textit{Traffic density} is commonly measured on a two-stage scale (congested/non-congested) \cite{Gatto:2020:MLTCD:IEEETTS} or on a three-stage scale as either low/free (equivalent to vehicle speeds larger equal to 40 km/h), medium (20-40 km/h), or heavy/jammed (below 20 km/h)  \cite{Warghade:2017:RAcoustics:ICCCUBEA, Tyagi:2012:VTDSE:IEEETTS, Borkar:2013:NFC:ICFS, Bhandarkar:2014:VMC:ICCICN}.
Traffic density can also be measured by \textit{detecting and counting passing vehicles}. A common approach is to investigate run-time differences between stereo audio signals \cite{Chen:2001:DSFM:IEVT, Barbagli:2012:ASNVTM:ICAVSTA, Ishida:2016:DTW:ITSWC}.
Moving sources exhibit a sweep-like peak contour in the temporal development of the cross-correlation function between both signals.
While the contour's diagonal alignment indicates the direction of movement, the contour angle indicates the speed of a vehicle. Ishida et al. match pre-defined templates with the cross-correlation function to detect for left-right and right-left movements \cite{Ishida:2016:DTW:ITSWC}. In contrast, our experiments shown in \secref{sec:experiments} indicate that a convolutional neural network can learn such characteristic patterns by itself.

Heavy traffic can lead to \textit{traffic accidents}, which can be detected by the two sounds tire skidding and car crash \cite{Foggia:2016:Traffic:IEEE, Li:2018:Traffic:IEEE}.
Another approach for traffic monitoring is to distinguish between vehicles in good and bad \textit{mechanical condition} based on emitted sounds \cite{Bhandarkar:2014:VMC:ICCICN}.

Different audio signal representations are used for solving ATM tasks. 
While most often raw spectrogram representations are used, some authors compute more advanced audio features such as the Mel-frequency Cepstral Coefficients (MFCC) prior to the modeling and classification steps \cite{Bhandarkar:2014:VMC:ICCICN, Warghade:2017:RAcoustics:ICCCUBEA}. In order to train more robust algorithms, Gatto et al. apply data augmentation and mix recorded audio signals with additional noise \cite{Gatto:2020:MLTCD:IEEETTS}.

ATM algorithms apply various mostly traditional classification algorithms such as Nearest Neighbor classifier \cite{Warghade:2017:RAcoustics:ICCCUBEA}, Bayer's classifier \cite{Tyagi:2012:VTDSE:IEEETTS}, Random Forest classifier \cite{Gatto:2020:MLTCD:IEEETTS},  Support Vector Machines (SVM) \cite{Bhandarkar:2014:VMC:ICCICN}, Artificial Neural Networks (ANN) \cite{Bhandarkar:2014:VMC:ICCICN, Warghade:2017:RAcoustics:ICCCUBEA}, as well as hybrid approaches such as the Neuro-Fuzzy Classifier \cite{Borkar:2013:NFC:ICFS}. While most above-mentioned tasks require classification algorithms, Djukanovi{\'c} et al. use Support Vector Regression (SVR) \cite{Djukanovic:2020:Traffic:IVS} to predict the vehicle-to-microphone distances.

Most publications for ATM systems rely on proprietary datasets. However, some publicly available datasets are applicable for traffic monitoring.
The MIVIA road audio events data set includes audio recordings of 400 sound events of the two classes tire skidding and car crashes  \cite{Foggia:2016:Traffic:IEEE}. 
The MAVD dataset \cite{Zinemanas:2019:MAVD:DCASE} was published for sound event detection (SED) of particular traffic sounds which were recorded from the vehicle classes car, truck, bus, motorcycle in different states such as idling, accelerating, or braking.
In the research field of SED, many datasets such as the FSK50k \cite{Fonseca:2020:FSD50K:ARXIV} or the AudioSet \cite{Gemmeke:2017:Audioset:ICASSP} include general sound classes such as car, truck, or train, whose recognition could be applied in ATM systems. Similarly, 
acoustic scene classification (ASC) datasets such as the TUT Urban Acoustic Scenes 2018 dataset \cite{Mesaros:2018:MultiDeviceDataset:DCASE} allow to train algorithms to detect amongst others traffic-related sound scenes such as ``Street, traffic'', ``Bus'', ``Metro'', and ``Tram''.






\section{IDMT-Traffic dataset}
\label{sec:dataset}

In this section, we introduce a novel dataset for acoustic traffic monitoring (\texttt{IDMT-Traffic})\footnote{The dataset will be made publicly available soon at \url{https://www.idmt.fraunhofer.de/en/publications/datasets.html}}.
It is intended as a public evaluation benchmark for the detection and classification of passing vehicles on inner-city and overland roads.
The dataset includes time-synchronized stereo audio recordings made with both high-quality sE8 microphones\footnote{\url{https://www.seelectronics.com/se8-mic}} as well as lower-budget microelectro-mechanical systems (MEMS) microphones\footnote{InvenSense ICS-43434} from four different recording locations including three city traffic locations and one country road location in and around Ilmenau, Germany.
The recording scenarios include different speed limits (30, 50, and 70 km/h) as well as wet and dry road conditions.

\figref{fig:setup_1} and \figref{fig:setup_2} illustrate the recording setup, which was placed with a distance of 0.5 meters to the adjacent street.
Both pairs of sE8 and MEMS microphones are fixed at a distance of 18.5 cm.
Video recordings were solely made for the purpose of annotating the type  and movement direction of passing vehicles. For reasons of data protection, we strictly avoided filming faces and licence plates by aiming the camera at the lower part of the vehicles as shown in \figref{fig:video1} and \figref{fig:video2}.
For each of both microphone types, around 2.5 hours of audio recordings exist with a total of 4718 annotated passing vehicles.
The dataset includes four classes: cars (3903 events), trucks (511 events), busses (53 events), and motorcycles (251 events). This distribution reflects the natural imbalance of vehicle types in common traffic scenarios. 

\begin{figure*}[t]
\begin{subfigure}{.24\textwidth}
  \centering
  \includegraphics[width=.7\linewidth]{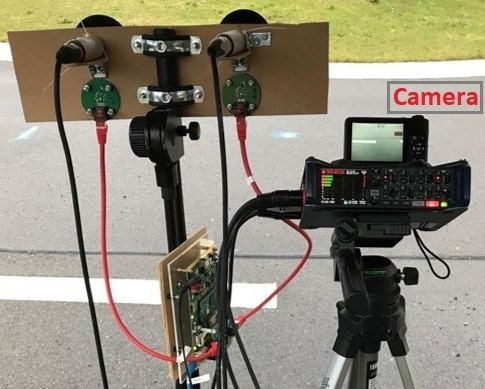}  
  \caption{Back view: stereo microphone setup (left) and video camera (right).}
  \label{fig:setup_1}
\end{subfigure}
\begin{subfigure}{.24\textwidth}
  \centering
  \includegraphics[width=.7\linewidth]{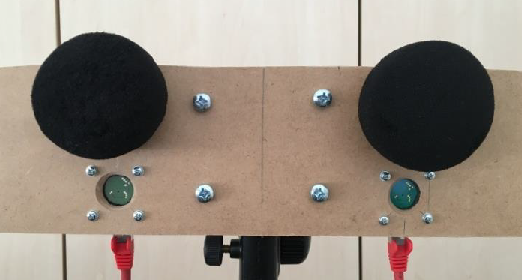}  
  \caption{Frontal view: microphone setup (top: sE8, bottom: MEMS microphones.)}
  \label{fig:setup_2}
\end{subfigure}
\begin{subfigure}{.24\textwidth}
 \centering
  \includegraphics[width=.7\linewidth]{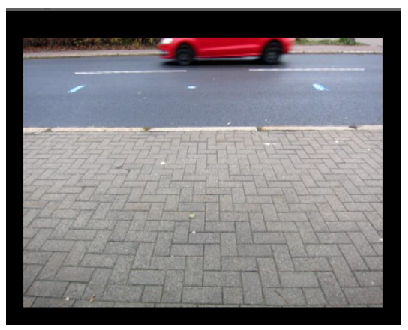}  
  \caption{Example video frames for vehicle type annotations.}
  \label{fig:video1}
  \label{fig:sub-first}
\end{subfigure}
\begin{subfigure}{.24\textwidth}
  \centering
  \includegraphics[width=.7\linewidth]{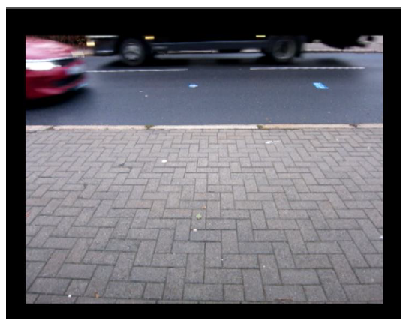}  
  \caption{Example video frames for vehicle type annotations.}
  \label{fig:video2}
\end{subfigure}
\caption{Audio-video recording setup for the dataset creation.}
\label{fig:recording_setup}
\end{figure*}

\section{Benchmark Experiments}
\label{sec:experiments}

Using the \texttt{IDMT-Traffic} dataset introduced in \secref{sec:dataset}, we conducted two benchmark experiments for different ATM tasks. Here, we only used the audio recordings recorded with the hiqh-quality sE8 microphones.

\begin{figure*}[t]
\begin{subfigure}{.23\textwidth}
  \centering
  \includegraphics[width=.8\linewidth]{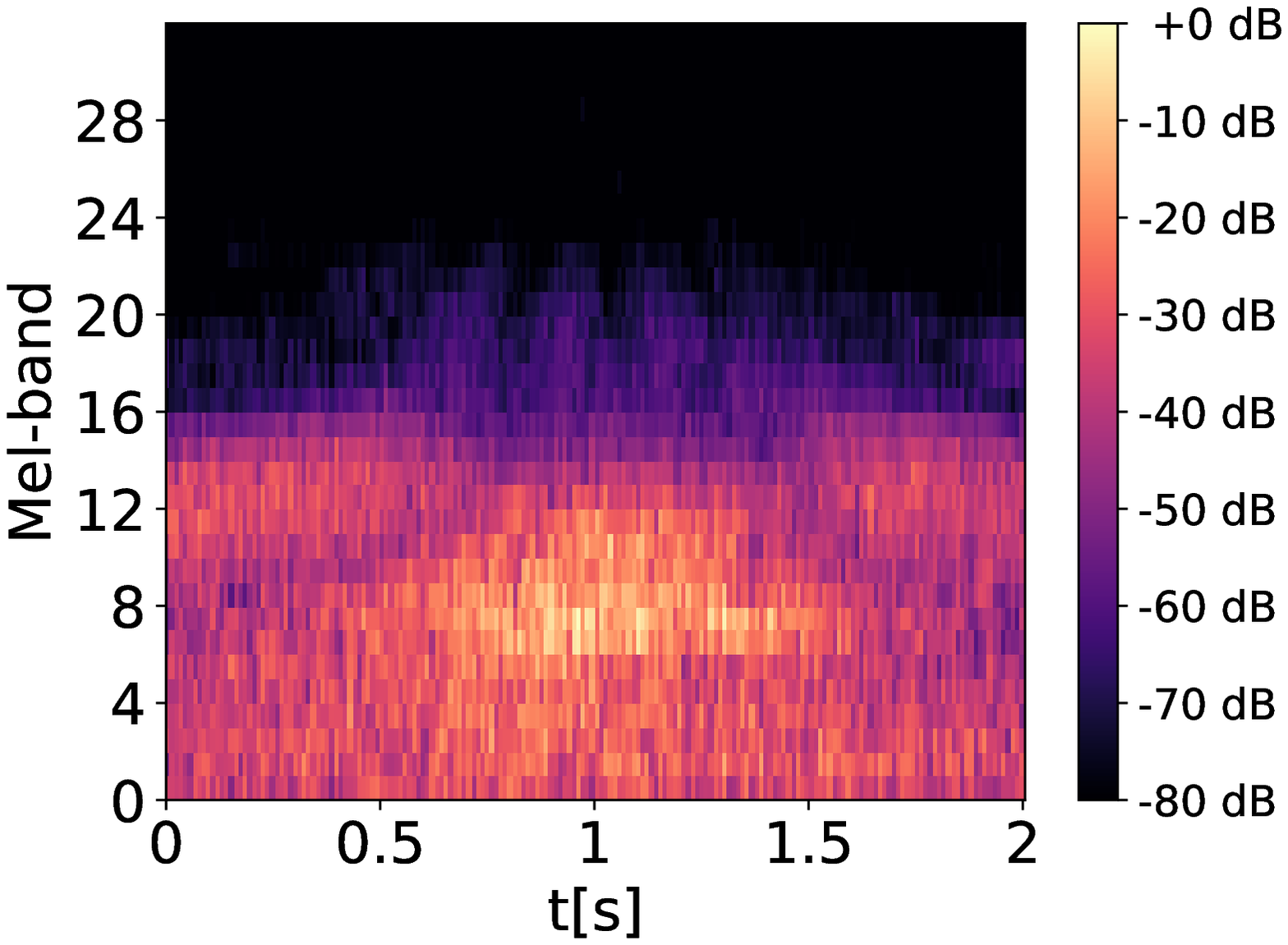}  
  \caption{CA, L$\rightarrow$R, 50 km/h, MS.}
  \label{fig:sub-first}
\end{subfigure}
\begin{subfigure}{.23\textwidth}
  \centering
  \includegraphics[width=.8\linewidth]{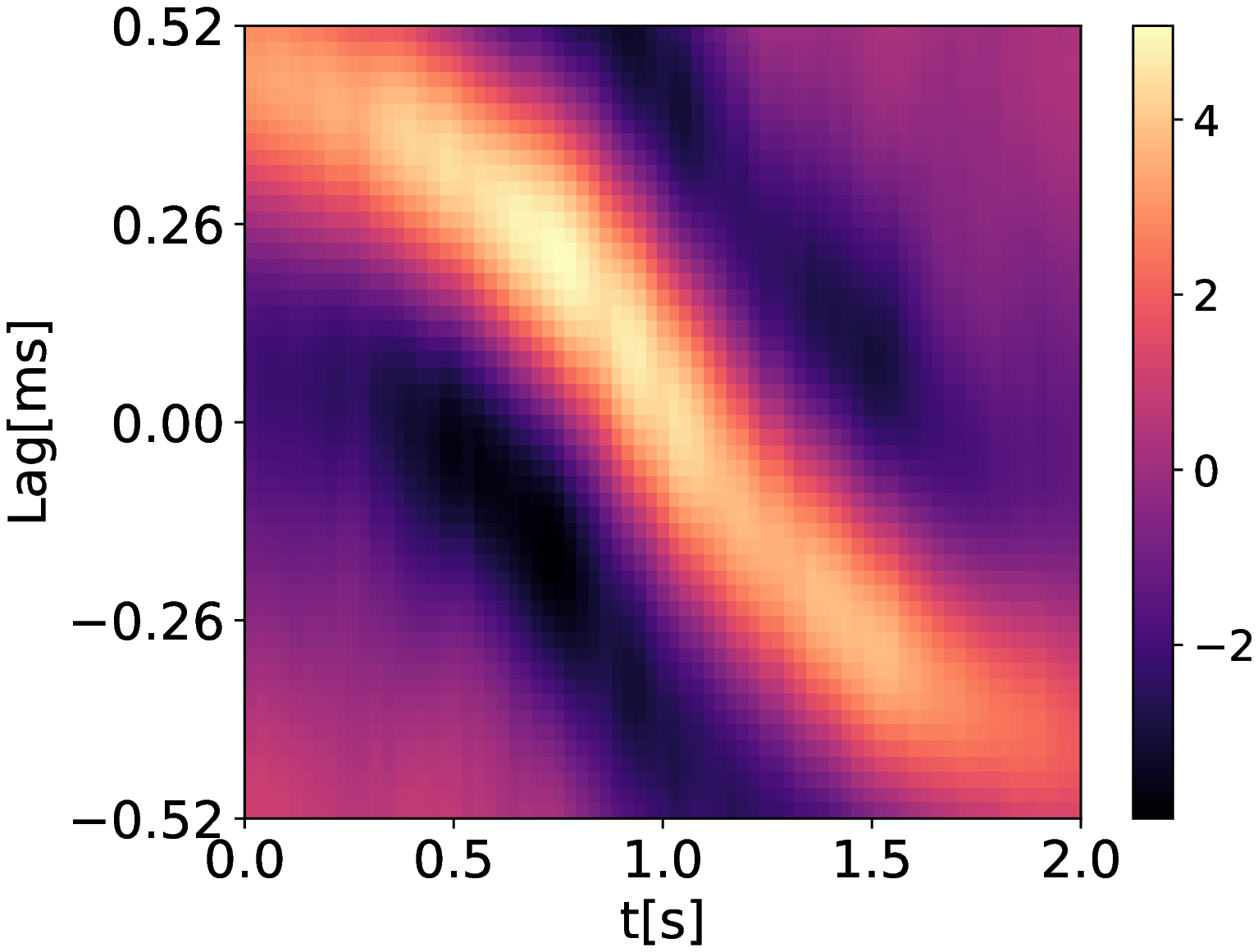}  
  \caption{CA, L$\rightarrow$R, 50 km/h, CC.}
  \label{fig:sub-second}
\end{subfigure}
\begin{subfigure}{.23\textwidth}
  \centering
  \includegraphics[width=.8\linewidth]{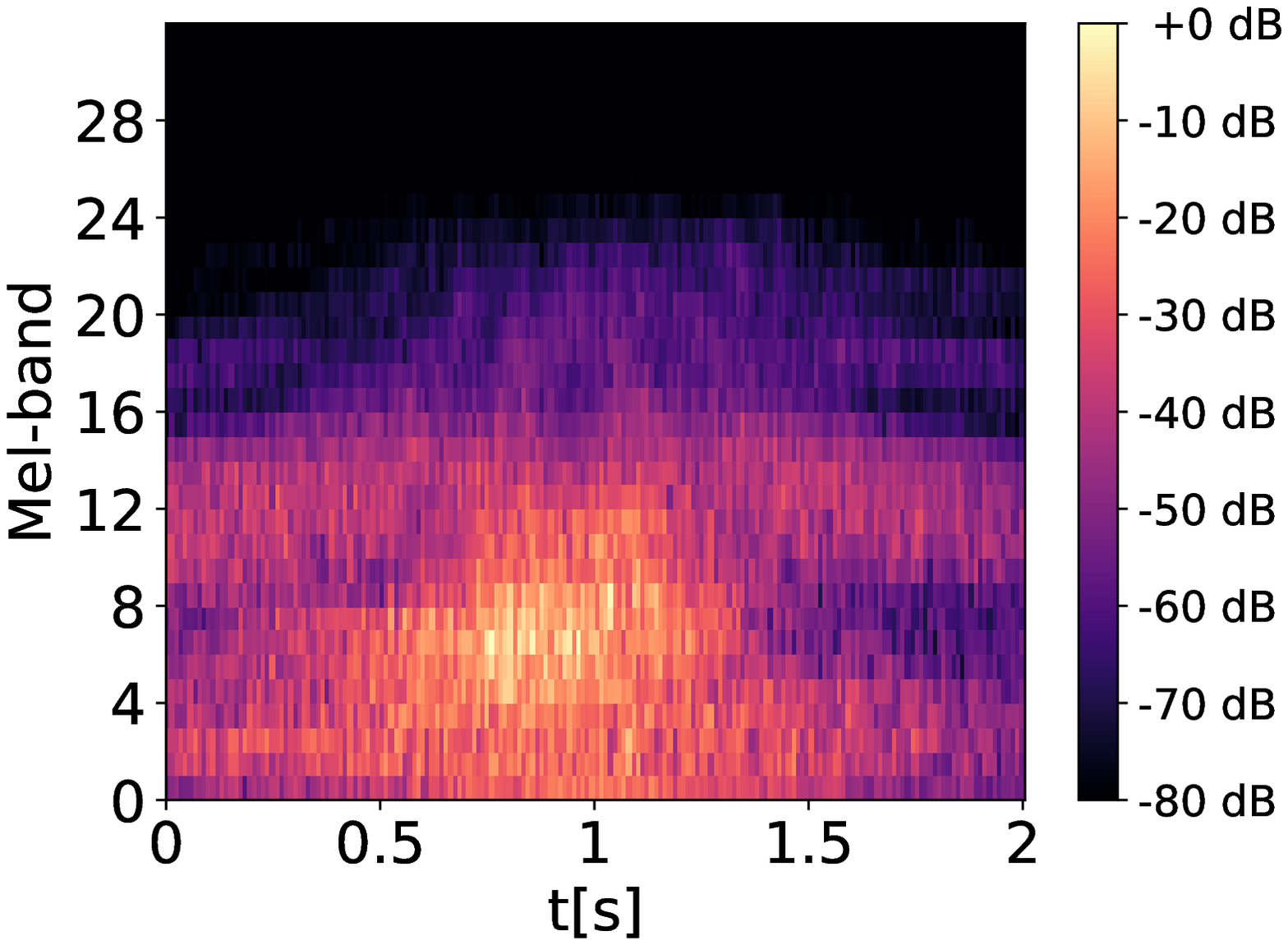}  
  \caption{TR, R$\rightarrow$L, 50 km/h, MS.}
  \label{fig:sub-first}
\end{subfigure}
\begin{subfigure}{.23\textwidth}
  \centering
  \includegraphics[width=.8\linewidth]{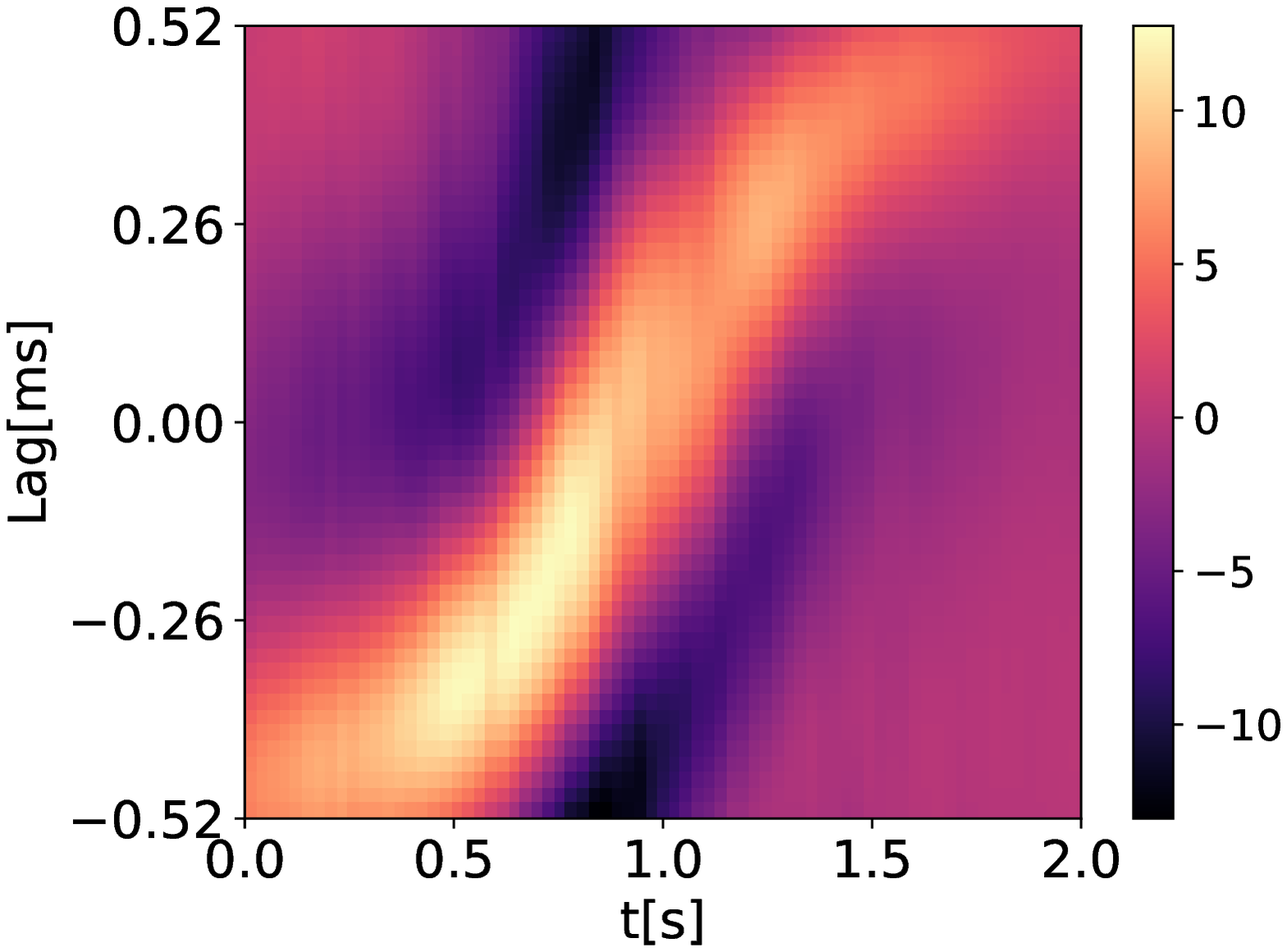}  
  \caption{TR, R$\rightarrow$L, 50 km/h, CC.}
  \label{fig:sub-second}
\end{subfigure}
\begin{subfigure}{.23\textwidth}
  \centering
  \includegraphics[width=.8\linewidth]{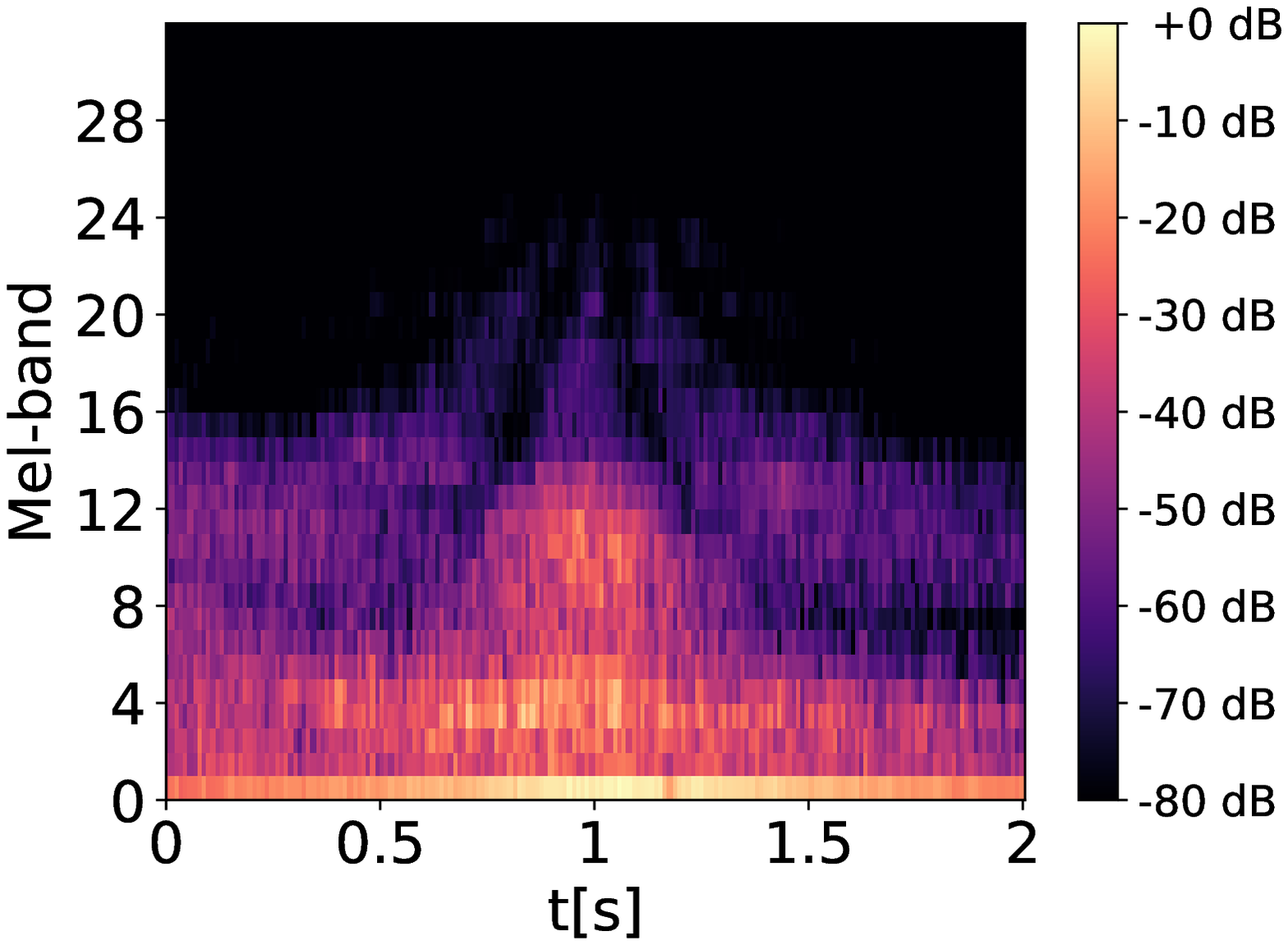}  
  \caption{MC, L$\rightarrow$R, $\approx$ 70 km/h, MS.}
  \label{fig:sub-first}
\end{subfigure}\hfill
\begin{subfigure}{.23\textwidth}
  \centering
  \includegraphics[width=.8\linewidth]{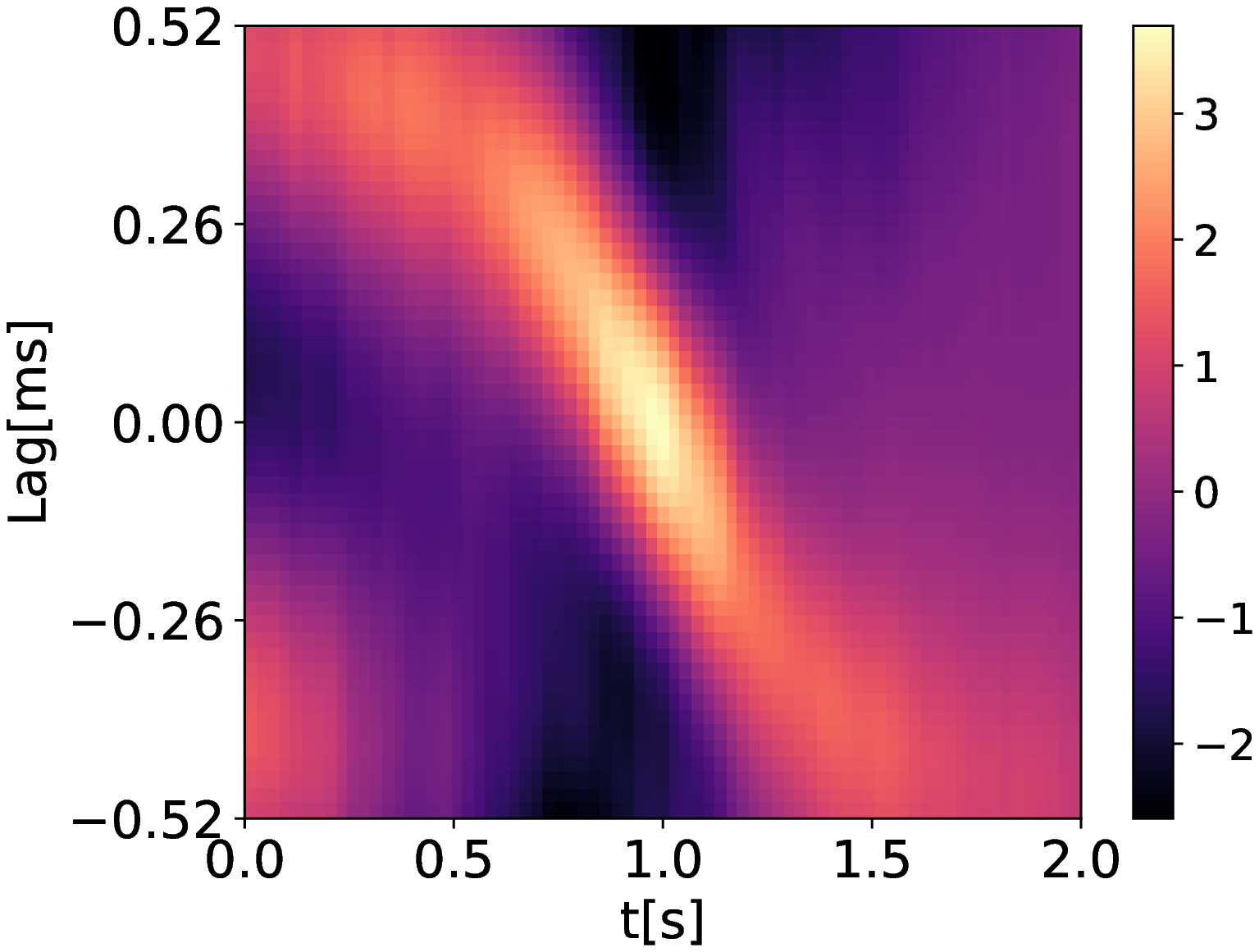}  
  \caption{MC, L$\rightarrow$R, $\approx$ 70 km/h, CC.}
  \label{fig:sub-second}
\end{subfigure}

\caption{Examples of two-second long patches taken from the \texttt{IDMT-Traffic} dataset for the vehicle type classes car (CA), truck (TR), and motorcycle (MC) for both mel-spectrogram (MS) and cross-correlation (CC) features. Direction of movement is either left-to-right (L$\rightarrow$R) or right-to-left (R$\rightarrow$L).}
\label{fig:examples}
\end{figure*}

\subsection{Audio Representation \& Pre-processing}
\label{sec:audio_proc}

As a first processing step, the left and right channels of the  stereo audio files were averaged to a mono channel at the original sample rate of 48 kHz. 
In this work, we address the tasks of detecting vehicles and classifying their type and direction of movement.
Therefore, we extract two types of features to be processed by the convolutional neural networks introduced in \secref{sec:models}.

As first feature type, we extract mel-spectrograms using the \textit{librosa} library \cite{McFee:2015:Librosa:SCIPY}.
We average the left and right audio channels and down-sample the audio signal to a sample rate of 22.05 kHz. 
Mel-spectrograms are computed using an FFT size, a window size, and a hop size of 2048, 1024, and 512 samples, respectively. 
In our experiments, we investigate the effect of changing the number of mel-bands $N_B\in\{16,32,64,128\}$ on the recognition performance.
Log-magnitude scaling is applied in order to compensate for the natural dynamic range of the traffic recordings. Finally, two-second long sub-sequences (patches) are extracted. 

As a second feature type, we compute the local cross-correlation between the left and right audio channel at the original sample rate of 48 kHz.
We extract blocks of 200 ms length with a hopsize of 25 ms from the audio signals.
From the cross-correlation function between the left and right channel of the $b$-th block, we keep the center part $c_b \in \mathbb{R}^{51}$ with a margin of 25 lags around zero-lag index.
We derive a two-dimensional feature representation by stacking such cross-correlation blocks for two-second long patches as for the mel-spectrograms.

Features are standardized (zero mean and unit variance) per bin, i.\,e. per frequency bin for the mel-spectrogram patches and per time lag for the cross-correlation patches, over all patches of a given dataset. This normalization procedure is performed independently for the training set, validation set, and test set.
\figref{fig:examples} shows both the mel-spectrogram as well as the cross-correlation features for three examples, including passings of a car, a truck, and a motorcycle.

\subsection{Neural Network Architectures}
\label{sec:models}

In our benchmark experiments, we test three different convolutional neural network architectures, which will be detailed in the following sections. \tabref{tab:params_per_model} summarizes the number of parameters per model.

\subsubsection{VGGNet}
\label{sec:vggnet}

The \texttt{VGGNet} model proposed by Takahashi et al. in \cite{Takahashi:2017:ASC:APSIPA} uses four pairs of 3x3 convolutional layers with intermediate pooling only between the layer pairs. This way, the spatial resolution is decreased while the number of filters is increased from 32 to 256.
Several regularization strategies such as batch normalization, dropout as well as L2 regularization in the penultimate dense layers are applied to improve the model's generalization towards new data.

\subsubsection{ResNet}
\label{sec:resnet}

The \texttt{ResNet} is the ``RN1'' as proposed by Koutini et al. in \cite{Koutini:2019:ResNet:EUSIPCO}. 
It includes five residual blocks with two convolutional layers each. The network was designed to have a reduced receptive field and has been evaluated the task of acoustic scene classification.

\subsubsection{SqueezeNet}
\label{sec:squeezenet}

The \texttt{SqueezeNet} architecture was introduced in \cite{Iandola:2016:SqueezeNet:ICLR} and implements several model compression strategies. As a first strategy, 3x3 filters are replaced by 1x1 filters in the convolutional layers. As a second strategy, the network includes a number fire modules, which use a squeeze-and-expand approach to reduce the depth of feature maps while maintaining their size.

\subsubsection{MobileNetMini}
\label{sec:mobile_net_mini}

The \texttt{MobileNetMini} model is a miniaturized version of the MobileNet architecture proposed in \cite{Sandler:2018:MobileNet:CVPR}.
It includes one convolutional layer and one depthwise convolutional layer with batch normalization and ReLU activation functions each followed by a global max pooling operations and a final softmax dense layer.

\begin{table}[t]
 \begin{center}
\scalebox{1}{  
\begin{tabular}{p{0.13\textwidth}p{0.1\textwidth}p{0.18\textwidth}}
 \toprule
 \textbf{Model} & \textbf{\# Parameters}  & \textbf{Classification Task}  \\
 \midrule
 \texttt{VGGNet} & $1,442,788$ & Vehicle type\\
 \texttt{ResNet} & $3,259,012$ & Vehicle type\\
 \texttt{SqueezeNet} & $1,171,652$ &  Vehicle type\\
 \texttt{MobileNetMini} & $15,363$ & Direction of movement\\
 
\bottomrule
 \end{tabular}
 }
\end{center}
  \caption{Summary of the compared neural network architectures, their number of parameters, as well as the classification tasks, they have been evaluated for.}
\label{tab:params_per_model}
\end{table}

\subsection{Experimental Procedure}
\label{sec:experimental_procedure}

From the \texttt{IDMT-Traffic} dataset, we first select audio files recorded at two locations with speed limits at 30 and 50 km/h.
Both sets are combined, then shuffled and split into training set (90 \%) and validation set (10 \%).
Recordings from the third location (speed limit at 70 km/h) was used as test set.
Using this data partition, we aim to test the robustness of the ATM algorithms against different vehicle speeds and the corresponding changes in the vehicle sound characteristics.
We trained all neural networks using the Adam optimizer \cite{Kingma:2015:Adam:ICLR} for 250 epochs with a learning rate of  $10^{-5}$. Early stopping with a patience of 50 epochs is used on the validation loss to monitor the training process.

\begin{table}[t]
 \begin{center}
\scalebox{1}{  
\begin{tabular}{p{0.1\textwidth}p{0.03\textwidth}p{0.03\textwidth}p{0.07\textwidth}p{0.1\textwidth}}
 \toprule
 \textbf{Dataset} & \textbf{Car}  & \textbf{Truck}  & \textbf{Motorcycle}  & \textbf{No vehicle}  \\
 \midrule 
  Training Set & $2471$ & $290$ &$132$ & $2393$ \\
  Validation Set & $275$ &$32$  &$15$  & $266$\\
  Test Set & $1157$ & $189$ & $99$ & $1412$ \\
\bottomrule
 \end{tabular}
 }
\end{center}
  \caption{Number of patches per class in the training set, validation set, and test set for vehicle type classification.}
\label{tab:dataset_patches}
\end{table}


\begin{table}[t]
 \begin{center}
\scalebox{1}{  
\begin{tabular}{p{0.1\textwidth}p{0.03\textwidth}p{0.03\textwidth}p{0.07\textwidth}p{0.1\textwidth}}
 \toprule
 \textbf{Model} & \textbf{Car}  & \textbf{Truck}  & \textbf{Motorcycle}  & \textbf{No vehicle}  \\
 \midrule 
 \multicolumn{4}{l}{$N_B=16$} \\
 \midrule 
  \texttt{VGGNet} & 0.94 & 0.5 & 0.96 & 1.0 \\
  \texttt{ResNet} & 0.94 & 0.49 & 0.96 & 1.0\\
  \texttt{SqueezeNet} & 0.92 & 0.48 & 0.9 & 1.0 \\
 \midrule 
 \multicolumn{4}{l}{$N_B=32$} \\
 \midrule 
  \texttt{VGGNet} & 0.94 & 0.46 & 0.96 & 1.0 \\
  \texttt{ResNet} & 0.94 & 0.44 & 0.97 & 1.0\\
  \texttt{SqueezeNet} & 0.94 & 0.42 & 0.9 & 1.0 \\
 \midrule 
 \multicolumn{4}{l}{$N_B=64$} \\
 \midrule 
  \texttt{VGGNet} & 0.94 & 0.49 & 0.97 & 1.0 \\
  \texttt{ResNet} & 0.94 & 0.49 & 0.97 & 1.0\\
  \texttt{SqueezeNet} & 0.94 & 0.5 & 0.95 & 1.0 \\
 \midrule 
 \multicolumn{4}{l}{$N_B=128$} \\
 \midrule 
  \texttt{VGGNet} & 0.94 & 0.44 & 0.96 & 1.0 \\
  \texttt{ResNet} & 0.94 & 0.45 & 0.95 & 1.0\\
  \texttt{SqueezeNet} & 0.91 & 0.53 & 0.97 & 1.0 \\
\bottomrule
 \end{tabular}
 }
\end{center}
  \caption{Class-wise f-score results for vehicle type classification using the three neural network models for mel-spectrograms with 16 bins.}
\label{tab:results_exp_1}
\end{table}

\subsection{Experiment 1 - Vehicle Type Classification}

We consider a four-class classification scenario where we include the three vehicle types cars, trucks, and motorcycles as well a no-vehicle class, which includes spectrogram patches without any passing vehicles.
The patches for the first three classes are centered around the annotated passing times. 
Non-vehicle patches were randomly sampled in between annotated vehicle passings in the audio recordings of the \texttt{IDMT-Traffic} dataset.
The number of patches per class as well as their partition to training set, validation set, and test set is given in \tabref{tab:dataset_patches}.
It can be observed that the classes no-vehicle and car have most patches followed by truck and motorcycle.

\tabref{tab:results_exp_1} summarizes the class-wise f-scores for the three investigated neural network architectures \texttt{VGGNet}, \texttt{ResNet}, and \texttt{SqueezeNet}.
We observe that all models perfectly recognize the no-vehicle patches therefore allow for a robust vehicle detection (binary classification task) based on the high-quality sE8 audio recordings.
Concerning the model performance, \texttt{VGGNet} and \texttt{ResNet} perform comparably well and slightly outperform the \texttt{SqueezeNet} model.
Interestingly, the results show that a frequency resolution of only 16 mel-bands ($N_B=16$) is sufficient to classify vehicle types.

\tabref{tab:conf_mat} illustrates as an example the confusion matrix for the \texttt{VGGNet} with $N_B=16$. It becomes apparent that the truck-to-car confusion is the most prominent misclassification. We assume that since both vehicles have only small differences in their geometric size, they cause a similar acoustic footprint which complicates their distinction.

\begin{table}[t]
 \begin{center}
\scalebox{1}{  
\begin{tabular}{p{0.1\textwidth}p{0.03\textwidth}p{0.03\textwidth}p{0.07\textwidth}p{0.1\textwidth}}
 \toprule
  & \textbf{Car}  & \textbf{Truck}  & \textbf{Motorcycle}  & \textbf{No vehicle}  \\
 \midrule 
\textbf{Car} & \textbf{97.29} & 2.62 & 0.02 & 0.09 \\
\textbf{Truck} & 60.21 & \textbf{38.84} & 0.63 & 0.32 \\
\textbf{Motorcycle} & 3.23 & 1.21 & \textbf{95.35} & 0.2\\
\textbf{No vehicle} & 0.23 & 0.01 & 0.11 & \textbf{99.65} \\
 \bottomrule
 \end{tabular}
 }
\end{center}
  \caption{Confusion matrix for vehicle type classification using \texttt{VGGNet} with $N_B=16$ (all values in percent).}
\label{tab:conf_mat}
\end{table}

\begin{table}[t]
 \begin{center}
\scalebox{1}{  
\begin{tabular}{p{0.1\textwidth}p{0.03\textwidth}p{0.03\textwidth}p{0.07\textwidth}p{0.1\textwidth}}
 \toprule
 \textbf{Dataset} & \textbf{L$\rightarrow$R}  & \textbf{R$\rightarrow$L}  & \textbf{No vehicle} \\
 \midrule 
  Training Set & $1445$ & $1448$ &$2393$ \\
  Validation Set & $161$ &$161$  &$266$\\
  Test Set & $678$ & $767$ & $1412$ \\
\bottomrule
 \end{tabular}
 }
\end{center}
  \caption{Number of patches per class in the training set, validation set, and test set for direction of movement estimation.}
\label{tab:dataset_patches_vmdd}
\end{table}

\begin{table}[t]
 \begin{center}
\scalebox{1}{  
\begin{tabular}{p{0.1\textwidth}p{0.05\textwidth}p{0.05\textwidth}p{0.1\textwidth}}
 \toprule
  & \textbf{L$\rightarrow$R}  & \textbf{R$\rightarrow$L}  &  \textbf{No vehicle}  \\
 \midrule 
\textbf{L$\rightarrow$R} & \textbf{96.61} & 0.83 & 2.57  \\
\textbf{R$\rightarrow$L} & 0.26 & \textbf{98.64} & 1.1  \\
\textbf{No vehicle} & 0 & 0.21 & \textbf{99.79} \\
 \bottomrule
 \end{tabular}
 }
\end{center}
  \caption{Confusion matrix for direction of movement estimation using the  \texttt{MobileNetMini} (all values in percent).}
\label{tab:conf_mat_2}
\end{table}

\subsection{Experiment 2 - Direction of Movement Estimation}

In this experiment, we evaluate the performance of the \texttt{MobileNetMini} architecture on the cross-correlation features for detecting the direction of movement. Here, we include patches across different vehicle types for the classes left-to-right and right-to-left and add no-vehicle patches as third class to simulate the detection task. The number of patches per class as well as their distribution among training, validation, and test sets is given in the \tabref{tab:dataset_patches_vmdd}.
As can be seen in the confusion matrix in \tabref{tab:conf_mat_2}, the direction of movement can be easily determined using a very small MobileNet architecture and the cross-correlation features. 
This result confirms the findings from the scientific literature \cite{Chen:2001:DSFM:IEVT, Barbagli:2012:ASNVTM:ICAVSTA, Ishida:2016:DTW:ITSWC}. As the only distinction, our method relies on automatic feature learning as part of the CNN model.

\section{Conclusions}
\label{sec:conclusion}

In this paper we show that acoustic traffic monitoring provides a low-cost and non-invasive alternative to traffic monitoring approaches based on other sensor modalities such as vision or radar. After providing a thorough review of scientific publications on acoustic traffic monitoring, we present the novel \texttt{IDMT-Traffic} dataset, which is a freely-accessible benchmark dataset intended to stimulate further research in acoustic traffic monitoring.

In our baseline experiments, which use solely the high-quality audio recordings in the dataset, we show that state-of-the-art convolutional neural networks already achieve high performance scores for vehicle type classification and direction of movement estimation. 
Furthermore, the results show that vehicle detection can be implemented easily with either the mel-spectrogram or the cross-correlation features.

Having the goal of a real-world deployment of an ATM system in mind, several challenges need to be addressed. The first challenge arises from the microphone mismatched between high-quality and low-quality microphones used in mobile sensor devices. By including audio recordings from both high-quality sE8 microphones as well as lower-quality MEMS microphones, the \texttt{IDMT-Traffic} dataset provides a suitable test-bed to develop new algorithmic strategies for domain adaptation.
A second challenge comes from computational performance constraints of mobile sensor devices, which might require to compress the neural network models.
In addition to these challenges, future research directions include a precise speed estimation of vehicles as well as an improved classification of passing trucks.

\section*{Acknowledgements}

This research was supported by 
the Fraunhofer Innovation Program ``Trusted Resource Aware ICT'' (TRAICT).
The authors would like to thank Ke Shen, Yishi Sun, Jiaming Tan, Julia Aileen, Katharina Duong, and Thanh Phong Duong for their assistance in the dataset recording process.



\bibliographystyle{IEEEtran}
\bibliography{refs_smt}

\begin{thebibliography}{10}
\providecommand{\url}[1]{#1}
\csname url@samestyle\endcsname
\providecommand{\newblock}{\relax}
\providecommand{\bibinfo}[2]{#2}
\providecommand{\BIBentrySTDinterwordspacing}{\spaceskip=0pt\relax}
\providecommand{\BIBentryALTinterwordstretchfactor}{4}
\providecommand{\BIBentryALTinterwordspacing}{\spaceskip=\fontdimen2\font plus
\BIBentryALTinterwordstretchfactor\fontdimen3\font minus
  \fontdimen4\font\relax}
\providecommand{\BIBforeignlanguage}[2]{{%
\expandafter\ifx\csname l@#1\endcsname\relax
\typeout{** WARNING: IEEEtran.bst: No hyphenation pattern has been}%
\typeout{** loaded for the language `#1'. Using the pattern for}%
\typeout{** the default language instead.}%
\else
\language=\csname l@#1\endcsname
\fi
#2}}
\providecommand{\BIBdecl}{\relax}
\BIBdecl

\bibitem{Borkar:2013:NFC:ICFS}
P.~Borkar and L.~G. Malik, ``{Acoustic Signal based Traffic Density State
  Estimation using Adaptive Neuro-Fuzzy classifier},'' in \emph{Proceedings of
  the IEEE International Conference on Fuzzy Systems (FUZZ-IEEE)}, Hyderabad,
  India, 2013, pp. 1--8.

\bibitem{Bhandarkar:2014:VMC:ICCICN}
M.~Bhandarkar and T.~Waykole, ``{Vehicular Mechanical Condition Determination
  and On Road Traffic Density Estimation Using Audio Signals},'' in
  \emph{Proceedings of the 6th International Conference on Computational
  Intelligence and Communication Networks}, Bhopal, India, 2014, pp. 395--401.

\bibitem{Warghade:2017:RAcoustics:ICCCUBEA}
V.~P. Warghade and M.~S. Deshpande, ``{Road Traffic Condition Estimation Based
  on Road Acoustics},'' in \emph{Proceedings of the International Conference on
  Computing, Communication, Control and Automation (ICCUBEA)}, Pune, India,
  2017, pp. 1--5.

\bibitem{Tyagi:2012:VTDSE:IEEETTS}
V.~Tyagi, S.~Kalyanaraman, and R.~Krishnapuram, ``{Vehicular Traffic Density
  State Estimation Based on Cumulative Road Acoustics},'' \emph{IEEE
  Transactions on Intelligent Transportation Systems}, vol.~13, no.~3, pp.
  1156--1166, 2012.

\bibitem{Gatto:2020:MLTCD:IEEETTS}
R.~C. Gatto and C.~H.~Q. Forster, ``{Audio-Based Machine Learning Model for
  Traffic Congestion Detection},'' \emph{IEEE Transactions on Intelligent
  Transportation Systems}, pp. 1--8, 2020.

\bibitem{Djukanovic:2020:Traffic:IVS}
S.~Djukanovic, J.~Matas, and T.~Virtanen, ``{Robust Audio-Based Vehicle
  Counting in Low-to-Moderate Traffic Flow},'' in \emph{Proceedings of the IEEE
  Intelligent Vehicles Symposium (IV)}, Las Vegas, NV, USA, 2020, pp.
  1608--1614.

\bibitem{Chen:2001:DSFM:IEVT}
{Shiping Chen}, {Ziping Sun}, and B.~Bridge, ``{Traffic Monitoring Using
  Digital Sound Field Mapping},'' \emph{IEEE Transactions on Vehicular
  Technology}, vol.~50, no.~6, pp. 1582--1589, 2001.

\bibitem{Na:2015:ADNI:IEEEIC}
Y.~Na, Y.~Guo, Q.~Fu, and Y.~Yan, ``{An Acoustic Traffic Monitoring System:
  Design and Implementation},'' in \emph{Proceedings of the IEEE
  UIC-ATC-ScalCom-CBDCom-IoP}, Guangzhou, China, 2015, pp. 119--126.

\bibitem{Barbagli:2012:ASNVTM:ICAVSTA}
B.~Barbagli, G.~Manes, and R.~Facchini, ``{Acoustic Sensor Network for Vehicle
  Trafﬁc Monitoring},'' in \emph{Proceedings of the International Conference
  on Advances in Vehicular Systems, Technologies and Applications (VEHICULAR)},
  Venice, Italy, 2012, pp. 1--6.

\bibitem{Ishida:2016:DTW:ITSWC}
S.~Ishida, S.~Liu, K.~Mimura, S.~Tagashira, and A.~Fukuda, ``{Design of
  acoustic vehicle count system using DTW},'' in \emph{Proceedings of the ITS
  World Congress}, Melbourne, Australia, 2016, pp. 1--10.

\bibitem{Foggia:2016:Traffic:IEEE}
P.~Foggia, N.~Petkov, A.~Saggese, N.~Strisciuglio, and M.~Vento, ``{Audio
  Surveillance of Roads: A System for Detecting Anomalous Sounds},'' \emph{IEEE
  Transactions on Intelligent Transportation Systems}, vol.~17, no.~1, pp.
  279--288, 2016.

\bibitem{Li:2018:Traffic:IEEE}
Y.~Li, X.~Li, Y.~Zhang, M.~Liu, and W.~Wang, ``{Anomalous Sound Detection Using
  Deep Audio Representation and a BLSTM Network for Audio Surveillance of
  Roads},'' \emph{IEEE Access}, vol.~6, pp. 58\,043--58\,055, 2018.

\bibitem{Zinemanas:2019:MAVD:DCASE}
P.~Zinemanas, P.~Cancela, and M.~Rocamora, ``{MAVD: A Dataset for Sound Event
  Detection in Urban Environments},'' in \emph{Proceedings of the Workshop on
  Detection and Classification of Acoustic Scenes and Events (DCASE)}, New
  York, NY, USA, 2019, pp. 263--267.

\bibitem{Fonseca:2020:FSD50K:ARXIV}
E.~Fonseca, S.~Member, X.~Favory, J.~Pons, F.~Font, and X.~Serra, ``{FSD50K: an
  Open Dataset of Human-labeled Sound Events},'' \emph{arXiv preprint
  arXiv:2010.00475}, vol.~14, no.~8, pp. 1--21, 2020.

\bibitem{Gemmeke:2017:Audioset:ICASSP}
J.~F. Gemmeke, D.~P.~W. Ellis, D.~Freedman, A.~Jansen, W.~Lawrence, R.~C.
  Moore, M.~Plakal, and M.~Ritter, ``{Audio Set: An Ontology and Human-Labeled
  Dataset for Audio Events},'' in \emph{Proceedings of the IEEE International
  Conference on Acoustics, Speech and Signal Processing (ICASSP)}, New Orleans,
  LA, USA, 2017, pp. 776--780.

\bibitem{Mesaros:2018:MultiDeviceDataset:DCASE}
A.~Mesaros, T.~Heittola, and {Tuomas Virtanen}, ``{A Multi-Device Dataset for
  Urban Acoustic Scene Classification},'' in \emph{Proceedings of the Detection
  and Classification of Acoustic Scenes and Events (DCASE)}, Surrey, UK, 2018,
  pp. 9--13.

\bibitem{McFee:2015:Librosa:SCIPY}
B.~McFee, C.~Raffel, D.~Liang, D.~P.~W. Ellis, M.~McVicar, E.~Battenberg, and
  O.~Nieto, ``{librosa: Audio and Music Signal Analysis in Python},'' in
  \emph{Proceedings of the Scientific Computing with Python conference
  (Scipy)}, Austin, Texas, 2015, pp. 18--24.

\bibitem{Takahashi:2017:ASC:APSIPA}
G.~Takahashi, T.~Yamada, N.~Ono, and S.~Makino, ``{Performance Evaluation of
  Acoustic Scene Classification using DNN-GMM and Frame-Concatenated Acoustic
  Features},'' in \emph{Proceedings of the 9th Asia-Pacific Signal and
  Information Processing Association Annual Summit and Conference (APSIPA)},
  Honolulu, Hawaii, USA, 2018, pp. 1739--1743.

\bibitem{Koutini:2019:ResNet:EUSIPCO}
K.~Koutini, H.~Eghbal-Zadeh, M.~Dorfer, and G.~Widmer, ``{The receptive field
  as a regularizer in deep convolutional neural networks for acoustic scene
  classification},'' \emph{European Signal Processing Conference}, vol.
  2019-Septe, 2019.

\bibitem{Iandola:2016:SqueezeNet:ICLR}
\BIBentryALTinterwordspacing
F.~N. Iandola, S.~Han, M.~W. Moskewicz, K.~Ashraf, W.~J. Dally, and K.~Keutzer,
  ``{SqueezeNet: AlexNet-level accuracy with 50x fewer parameters and
  {\textless}0.5MB model size},'' in \emph{Proceedings of the International
  Conference on Learning Representations (ICLR)}, Toulon, France, 2016, pp.
  1--13. [Online]. Available: \url{http://arxiv.org/abs/1602.07360}
\BIBentrySTDinterwordspacing

\bibitem{Sandler:2018:MobileNet:CVPR}
M.~Sandler, A.~Howard, M.~Zhu, A.~Zhmoginov, and L.~C. Chen, ``{MobileNetV2:
  Inverted Residuals and Linear Bottlenecks},'' in \emph{Proceedings of the
  IEEE Computer Society Conference on Computer Vision and Pattern Recognition
  (CVPR)}, Salt Lake City, UT, USA, 2018, pp. 4510--4520.

\bibitem{Kingma:2015:Adam:ICLR}
D.~P. Kingma and J.~L. Ba, ``{Adam: A method for stochastic optimization},'' in
  \emph{Proceedings of the International Conference on Learning Representations
  (ICLR)}, San Diego, CA, USA, 2015, pp. 1--15.

\end{thebibliography}
%

\end{document}